\documentstyle[12pt]{article}
\begin{document}

\begin{titlepage}

\hfill{July 1994}

\hfill{}

\hfill{IP-ASTP-14}

\vskip 1 cm

\centerline{{\bf \large A novel left-right symmetric model}}

\vskip 1.5 cm

\centerline{{\large R. Foot}$^{(a)}$\footnote{
Email: Foot@hep.physics.mcgill.ca}
{\large and H. Lew}$^{(b)}$}

\vskip 1.0 cm
\noindent
\centerline{{\it (a) Department of Physics, McGill University,}}
\centerline{{\it 3600 University Street, Montreal, Quebec, Canada.}}

\vskip 0.25cm
\noindent
\centerline{{\it (b) Institute of Physics, Academia Sinica,}}
\centerline{{\it Nankang, Taipei, Taiwan 11529, R.O.C.}}

\vskip 1.0cm

\centerline{Abstract}
\vskip 1cm
\noindent
A novel gauge model which has a spontaneously broken parity
symmetry is constructed. The model has only 2 parameters beyond
those of the standard model. Some of the unusual implications of
the model are discussed.

\end{titlepage}

The standard model of particle physics (SM) is an extremely successful
description of past and existing experiments. At present there is no
experimental indication of a deficiency in the SM. In view of this it
is not surprising that there is divided opinion about what lies
beyond the SM. However, a diversity of opinions on how to connect the
known to the unknown is not necessarily a bad thing given that there
exists no guaranteed way of making this connection. On the other hand,
a reading of the current literature on particle physics could give the
mistaken view that the ideas of grand unification and supersymmetry
are so well motivated and unique that it is only a matter of time before
they are found to be true. Indeed, some people are even advocating
spending billions of dollars to test some of the parameter space
of some supersymmetric grand unified models \cite{sus}. We feel that
this attitude to physics beyond the SM is an extremely biased interpretation
and extrapolation of existing knowledge. Unfortunately, this bias in the
literature has remained unchecked and has been (and still is)
detrimental to the advance of particle physics (and to the tax payer)
in terms of both theory and experiment. It is detrimental to theoretical
physics in the sense that the number and variety of
hypotheses generated to guess at the new physics becomes suppressed.
The situation for experimental physics is no better because
the bias in theory leads to a bias in experiments.
Finally, it is tax payer unfriendly because the proponents of
supersymmetric grand unified theories demand large
accelerators to be built which are currently very expensive.

The current situation is quite different to the era before the
discovery of the W and Z gauge bosons.  The theoretical
case for the existence of the W and Z gauge bosons
was very strong. Their masses could be approximately predicted from
the data already obtained in low energy experiments.
In this case there were strong physics
reasons to build the necessary colliders to study these
gauge bosons. Unfortunately, the standard model works so well,
that there is, at present, no experimental evidence
for new physics beyond the standard model. Of course, this does
not mean that there is no new physics beyond the standard model.
In all likelihood there is new physics, but there is
an infinite number of possibilities for what this new physics might be.
In light of this current situation, it is rather unimaginative
of the particle physics community to spend so much effort repeatedly
studying the possibility of supersymmetric grand unified theories.
This does little to advance the theory of elementary particles.
In our opinion, these theories are uninteresting.
Grand unification can be motivated from experiment because they can
simplify the gauge quantum numbers of the quarks and leptons but they
are uninteresting because they involve new physics at untestable energy
scales \cite{guts}. Supersymmetry on the other hand is perhaps testable
but is uninteresting because it is not well motivated by experiment.
{\it It doesn't explain the existence of any
known particle or symmetry.}

In view of the above, we feel that it is important to search
for interesting new ideas for new physics beyond the SM (rather
than work on the same boring idea over and over again).
If particle physics is to advance new ideas are clearly needed.
One approach that we have been studying is the issue of the origin
of parity violation in nature. Parity is a natural candidate (in view
of one's experience with classical physics) for a symmetry in particle
physics. The fact that experiments have established the V-A nature of
the weak interactions does not mean that parity cannot be a symmetry of
nature. The point is that there is no unique definition of the parity
transformation in quantum field theory. Parity, by definition, takes
$x$ to $-x$ and it also transforms left-handed fermion fields to
right-handed fermions fields, but since there are many left- and
right-handed fermion fields, there is no unique definition.
However, one can check that the SM has no definition
of parity which can be a symmetry of the SM Lagrangian.
Thus, if parity is a symmetry of the Lagrangian, new physics must exist.
The form of this new physics is not uniquely determined however.

One popular way for parity to be a symmetry of the Lagrangian
is to extend the SM to include a $SU(2)_R$ symmetry.
This possibility has been extremely
well studied in the literature \cite{lr}.
This is but one possible realization of parity symmetry.
Another possibility is the quark-lepton symmetric model in which
the gauge group has a $SU(3)$ for leptons so that the gauge symmetry
is $SU(3) \otimes SU(3) \otimes SU(2)_L \otimes U(1)$ \cite{ql}.
This model has the parity operation interchanging the two $SU(3)$'s
and the left-handed (right-handed) quarks with the right-handed
(left-handed) leptons.
The purpose of the present paper is to
introduce one extremely economical
model which features spontaneously broken parity symmetry.

Before we introduce the model,
it will be helpful to discuss another,
but related, type of parity conserving model.
It is possible to define
a model in which parity is not broken at all:
neither explicitly nor spontaneously \cite{ep}. This is
achieved by doubling the fermion and boson
content of the standard model
(although the number of particles is doubled,
it turns out that the number of parameters
increases by only 2, making it the
simplest extension to the standard model
that we are aware of \cite{sob}).

Let us start with the standard model Lagrangian ${\cal L}_1$.
This Lagrangian is not invariant under the usual parity
transformation so it seems parity is violated. However,
this Lagrangian may not be complete.
If we add to ${\cal L}_1$ a new Lagrangian ${\cal L}_2$
which is just like ${\cal L}_1$ except that all left-handed
(right-handed) fermions are replaced by new right-handed
(left-handed) fermions which feel new interactions of the same
form and strength, then the theory described by
${\cal L} = {\cal L}_1 + {\cal L}_2$ is invariant under
a parity symmetry (under this symmetry ${\cal L}_1
\leftrightarrow {\cal L}_2$). In addition to these Lagrangian terms,
there may also be parity invariant terms which mix ordinary matter
with mirror matter.
We label this part of the Lagrangian as ${\cal L}_{int}$.
The terms in ${\cal L}_{int}$ are very important since they lead
to interactions between ordinary and mirror matter and hence
allow the idea to be experimentally
tested in the laboratory. The terms in
${\cal L}_{int}$ also contains the only new parameters of the model.
It turns out that there are only two
possible terms in ${\cal L}_{int}$ which
are gauge invariant, parity invariant and renormalizable.
For this reason the model has only 2
new parameters beyond those of the SM.

The gauge symmetry of the theory is
\begin{equation}
SU(3)_1 \otimes SU(2)_1 \otimes U(1)_1 \otimes
SU(3)_2 \otimes SU(2)_2 \otimes U(1)_2.
\label{gsym}
\end{equation}
There are two sets of fermions, the ordinary particles
(which we denote below by lower case script)
and their mirror images - the mirror particles
(which we denote below with uppercase script). The fields
transform under the gauge group of Eq.(\ref{gsym}) as
\begin{equation}
\begin{array}{ll}
f_L \sim (1, 2, -1)(1, 1, 0),& F_R \sim (1, 1, 0)(1, 2, -1),\\
e_R \sim (1, 1, -2)(1, 1, 0),& E_L \sim (1, 1, 0)(1, 1, -2),\\
q_L \sim (3, 2, 1/3)(1, 1, 0),&Q_R \sim (1, 1, 0)(3, 2, 1/3),\\
u_R \sim (3, 1, 4/3)(1, 1, 0),&U_L \sim (1, 1, 0)(3, 1, 4/3),\\
d_R \sim (3, 1, -2/3)(1, 1, 0),&D_L \sim (1, 1, 0)(3, 1, -2/3),
\end{array}
\end{equation}
(with generation index suppressed). The Lagrangian is invariant
under the discrete $Z_2$ parity symmetry defined by
\begin{equation}
\begin{array}{c}
x \rightarrow -x,\  t \rightarrow t,\\
G_1^{\mu} \leftrightarrow G_{2\mu},\
W_1^{\mu} \leftrightarrow W_{2\mu},\ B_1^{\mu}
\leftrightarrow B_{2\mu},\\
f_L \leftrightarrow \gamma_0 F_R,\ e_R \leftrightarrow
\gamma_0 E_L,\ q_L \leftrightarrow \gamma_0 Q_R,\
u_R \leftrightarrow \gamma_0 U_L,\ d_R \leftrightarrow
\gamma_0 D_L, \end{array}
\end{equation}
where $G_1^{\mu}(G_2^{\mu})$, $W_1^{\mu} (W_2^{\mu}) $
and $B_1^{\mu} (B_2^{\mu})$ are the gauge bosons of
the $SU(3)_1$ $(SU(3)_2)$,
$SU(2)_1 (SU(2)_2)$ and  $U(1)_1 (U(1)_2)$
gauge forces respectively.
The minimal model contains two Higgs doublets which are
also parity partners:
\begin{equation}
\phi_1 \sim (1, 2, 1)(1, 1, 0),\  \phi_2 \sim (1, 1, 0)(1, 2, 1).
\end{equation}

An important feature which distinguishes
this parity conserving theory
from other such theories (e.g., the usual left-right symmetric
model \cite{lr}) is that the parity symmetry is assumed
to be unbroken by the vacuum.
The most general renormalizable
Higgs potential can be written in the form
\begin{equation}
V(\phi_1, \phi_2) = \lambda_1 (\phi_1^{\dagger} \phi_1
+ \phi_2^{\dagger} \phi_2 - 2u^2)^2 + \lambda_2 (
\phi_1^{\dagger} \phi_1 - \phi_2^{\dagger} \phi_2)^2,
\label{pot1}
\end{equation}
where $\lambda_{1,2}$ and $u$ are arbitrary constants. In
the region of parameter space where $\lambda_{1,2} > 0$,
$V(\phi_1, \phi_2)$ is non-negative and is minimized
by the vacuum
\begin{equation}
\langle \phi_1 \rangle = \langle \phi_2 \rangle =
\left( \begin{array}{c}
0\\
u
\end{array}\right).
\end{equation}
The vacuum values of both Higgs fields are exactly the same
(provided $\lambda_{1,2} > 0$)
and hence parity is not broken by the vacuum in this theory.

If the solar system is dominated by the usual particles,
then the theory agrees with present experiments.
The idea can be tested in the laboratory because it is possible
for the two sectors to interact with each other via the terms
in ${\cal L}_{int}$.
In the simplest case that we are considering at the moment
(where ${\cal L}_1$ is the minimal SM lagrangian), there are just
two possible terms (i.e.,  gauge and parity invariant
and renormalizable) in ${\cal L}_{int}$. They are,
\vskip .3cm
\noindent
(1) The Higgs potential terms $\lambda \phi_1^{\dagger} \phi_1
\phi_2^{\dagger} \phi_2$ contained within  Eq.(\ref{pot1}) and
\vskip .3cm
\noindent
(2) The gauge boson kinetic mixing term $\omega
F^1_{\mu \nu} F^{2 \mu \nu}$
where $F^{1,2}_{\mu \nu} = \partial_{\mu} B^{1,2}_{\nu} -
\partial_{\nu} B^{1,2}_{\mu}$ (recall that $B_{\mu}^{1,2}$ are the
gauge bosons of $U(1)_{1,2}$ respectively).
\vskip .3cm
\noindent
The main phenomenological effect of the term in (1) is to modify
the interactions of the Higgs boson. This effect will be tested
if or when the Higgs scalar is discovered. The details have been
discussed in Ref.\cite{ep}. The main phenomenological effect of the
kinetic mixing term in (2) is to give small electric charges to
the mirror partners of the ordinary charged fermions. This effect
has also been discussed previously \cite{ep,mcp}.

Note that this parity symmetric theory has just 2 extra parameters
beyond the 20 parameters of the minimal standard model \cite{par}.
It is, as far as we are aware, the simplest
(in terms of parameter counting) known alternative to the SM.

Given the simplicity of this model,
it is interesting to look for other
similar models. In particular, if
we study the Higgs potential in
Eq.(\ref{pot1}), then there are just
two possible vacua. For the region
of parameter space with $\lambda_{1,2} > 0$ the vacuum
expectation values
(VEVs) of $\phi_1$ and $\phi_2$
are equal and parity is unbroken,
while in the region of parameter space where
$\lambda_1 + \lambda_2 > 0, and\   \lambda_2 < 0$,
one VEV is non-zero and the other is zero.
In other words parity is
broken spontaneously. It is this
alternative possibility that we will
discuss in this note. Observe that
this model has the same number of
parameters as the parity conserving model
(it has the same Lagrangian,
just a different range of parameters).

To facilitate discussion, we can rewrite the Higgs potential in terms
of the parameters $\lambda_1' = \lambda_1 + \lambda_2,  \  \lambda_2' =
- 4\lambda_2$:
\begin{equation}
V(\phi_1, \phi_2) = \lambda_1' (\phi_1^{\dagger} \phi_1 +
\phi_2^{\dagger} \phi_2 - u^2)^2 +  \lambda_2' (\phi_1^{\dagger}\phi_1
\phi_2^{\dagger}\phi_2).
\label{pot2}
\end{equation}
Written in this way, for $\lambda_{1,2}' > 0,$ the vacuum
\begin{equation}
\langle \phi_1 \rangle = u, \quad \langle \phi_2 \rangle = 0,
\label{vac1} \end{equation}
or
\begin{equation}
\langle \phi_1 \rangle = 0, \quad \langle \phi_2 \rangle = u,
\label{vac2}
\end{equation}
is manifest since the Higgs potential is non-negative
and equal to zero
for these two vacua.  The two vacua are degenerate, and we assume
that we live in a region of space described by
the vacuum Eq.(\ref{vac1}).
Note that the model has 5 physical scalar particles:
one neutral scalar with $m_{\phi^0_1}^2 = 4\lambda_1' u^2$,
two neutral scalars with $m_{\phi^0_2}^2 = \lambda_2' u^2$
and two (mini-) charged scalars with $m_{\phi^\pm_2}^2
= \lambda_2' u^2$.

At first sight it looks like the
mirror fermions and gauge bosons all have
zero masses because they all couple to
$\phi_2$ which has zero VEV.
However, dynamical effects of mirror
QCD condensation will induce a small mass for
the mirror $W$, $Z$ bosons as well as a
tiny VEV for $\phi_2$, and hence,
tiny masses for the mirror fermions.
Quantitatively, the Higgs potential
of Eq.(\ref{pot2}) gets modified by
\begin{eqnarray}
\Delta V & = & \sum_Q \lambda_Q \langle \bar Q_L Q_R \rangle \phi_2^0
+ \sum_q \lambda_q \langle \bar q_L q_R \rangle \phi_1^0
+ \hbox{\rm H.c.},
\nonumber \\
 & = & 2\sum_Q |\lambda_Q| \Lambda_m^3 Re(\phi_2^0)
+ 2\sum_q |\lambda_q| \Lambda^3 Re(\phi_1^0), \nonumber\\
 & \simeq & 2|\lambda_t| \Lambda_m^3 Re(\phi_2^0)
+ 2|\lambda_t| \Lambda^3 Re(\phi_1^0),
\end{eqnarray}
where $\Lambda_m^3 \equiv \langle \bar Q_L Q_R \rangle$,
$\Lambda^3 \equiv \langle \bar q_L q_R \rangle$ and the last line
comes from the fact that the top-quark contributions are the
largest (note that $\lambda_Q = \lambda_q$ because of parity symmetry).
As usual we will assume that the QCD condensates, $\Lambda_m$ and
$\Lambda$ are primarily determined by the light coloured particles
of the theory. However, $\Lambda_m$ and $\Lambda$ are not calculable,
at least not perturbatively, but we can still get some idea of the
relationship between $\Lambda_m$ and $\Lambda$ by considering the
running of the strong coupling parameters, $\alpha_s^m$ and
$\alpha_s$. At a scale above the electroweak symmetry breaking
scale, the renormalized coupling parameters
$\alpha_s^m$ and $\alpha_s$ will be equal and will
evolve in the same way until the top-quark threshold. Below this
threshold, the number of ``light'' mirror and ordinary quarks
are $n_F = 6$ and $n_f = 5$ respectively. Ignoring all other
quark thresholds we can naively extrapolate down to $\Lambda_m$
and $\Lambda$ using the one-loop approximation:
\begin{equation}
{\Lambda_m \over \Lambda} =
{\exp \left[ {-6\pi \over (33-2n_F)\alpha_s^m(m_t^2)} \right]
/ \exp \left[ {-6\pi \over (33-2n_f)\alpha_s(m_t^2)} \right]}.
\end{equation}
This gives $\Lambda_m / \Lambda \sim 0.5$
( with $\Lambda \sim 200$ MeV ).
The main point to draw from this is that $\Lambda_m < \Lambda$.
This is because there are more light mirror quarks than ordinary
ones and so the running of the strong coupling parameter is slower
in the mirror sector.

The Higgs potential with the QCD effects included can now be
minimized:
\begin{eqnarray}
 & \langle \phi_1^0 \rangle \simeq u, \quad
\langle \phi_2^0 \rangle
\simeq {|\lambda_t|\Lambda_m^3 / m_{\phi_2^0}^2}.
\label{vacp}
\end{eqnarray}
Note that the VEV of $\phi_2$ is expected to be very tiny
unless $m_{\phi_2^0}$ is very light.

The particle content in the mirror sector can be summarized
as follows:
\begin{itemize}
\item the mirror $W$ and $Z$ bosons will have masses of order
$\Lambda_m \sim 100$ MeV,
\item the four physical mirror
scalars will have mass $\sqrt{\lambda_2'}\ u$,
\item and the mirror fermions will have masses such that
\[
m_F = k m_f, \qquad \hbox{where} \qquad
k \equiv {\langle \phi_2 \rangle \over \langle \phi_1 \rangle }
\sim {g^2\over 2}{m_t \Lambda_m^3\over m_W^2 m_{\phi_2^0}^2}
\]
where $g$ is the $SU(2)$ gauge coupling
and $m_W$ is the usual $W$ boson
mass. By using $g^2 \sim 10^{-1}$,
$m_t \sim m_W  \sim 10^2$ GeV and
$\Lambda_m \sim 0.1$ GeV, gives
$k \sim 10^{-6}({\rm GeV}/m_{\phi_2^0})^2$.
For example, if $m_{\phi_2^0} \sim 10^2$ GeV,
then $k \sim 10^{-10}$.
This means that the masses of the mirror fermions can range from
about $10$ eV for the mirror top-quark to $10^{-4}$ eV for the
mirror electron. As this example illustrates, we expect $k$ to be
quite small, however strictly $m_{\phi_2^0}$ is a free parameter
of the theory, so that the value of $k$ cannot be predicted.
\end{itemize}

As in the exact parity model, the kinetic $\left[U(1)\right]^2$
gauge boson mixing term will induce a small
electric charge for the mirror partners of the charged particles of
the standard model. However, unlike the case where the parity symmetry
is unbroken, the masses of the mirror fermions will be smaller by the
constant factor $k$ (or $\langle \phi_2 \rangle / \langle \phi_1 \rangle$).
Once a light mini-charged particle is found one can check whether its mass
is consistent with a mirror partner of one of the known particles. If so,
then this will lead to a measurement of $k$ and hence the masses of all the
other mirror particles will be predicted. Also, the measurement of the
electric charge of the mirror fermion will fix the kinetic mixing parameter
and therefore the electric charges of all the mirror fermions will be
determined.

Note that the
$\left(\phi_1^{\dagger} \phi_1 \phi_2^{\dagger} \phi_2\right)$
mixing term in the Higgs potential can cause novel effects.
In particular, the ordinary Higgs will decay dominantly
into $\phi_2, \phi_2$ pairs provided $\phi_2$ is lighter than $\phi_1$
and lighter than the top-quark or gauge boson pair production threshold.
This decay mode will be nearly invisible because the mirror Higgs will
subsequently decay into mirror top-quark pairs and mirror $W$, $Z$ pairs.

Finally, we would like to make some cosmological comments.
Observe that this model cannot account for the observed
ratio of light element abundances in the early universe
(within the context of the standard big bang model of cosmology).
This is because the model has too many relativistic species,
(in addition to the usual light particles, the model has
mirror fermions and a mirror photon, contributing to the
energy density at the time of nucleosynthesis).
Within the standard big bang model, the large number
of relativistic species would cause the universe to expand
more rapidly and the weak interactions would freeze out
earlier resulting in a higher neutron to proton ratio,
and hence, a greater than observed Helium abundance.
This conclusion assumes that the mirror sector has the same
temperature as the ordinary sector (as well as the standard
assumptions of the big bang cosmology model). Observe that if a
temperature difference was set up between the ordinary and
mirror worlds at very high temperatures due to some unknown
mechanism, then this temperature difference would be washed out
due to the interactions between the ordinary and mirror worlds,
unless these interactions were sufficiently weak. To conclude
these cosmological comments, we also mention that this model
has a potential domain wall problem because of the discrete
parity symmetry being spontaneously broken \cite{dw}.

The inability of the model to account for the light element abundances
in the universe or resolve the potential domain wall problem does not
``rule out'' this or any other particle physics model. We believe that
only experiments can rule out (or confirm) a model of particle physics.
The reason that the standard cosmology model cannot rule out
models of particle physics (in our opinion) is because the cosmology model
is based on assumptions which are untested. It applies physical laws
outside their tested domain of validity. Also, it is incomplete in view of the
isotropic, homogeneity and flatness problems, to name a few \cite{kst}.
Our above comments are not intended to detract from
the important and interesting work on the standard big bang model.
Understanding the universe is a challenging problem, and the
standard big bang model is remarkably successful and consistent with the
standard model of particle physics, despite its incompleteness
and untested assumptions. However, having said this, we don't believe
that this success should be used to ``rule out'' models of particle
physics which happen to be less successful than the standard model
in describing the universe when used in conjuction with the standard
big bang model. Only experiments can ultimately determine whether a
model is ruled out or not.

In conclusion, we have explored the possibility of a spontaneously
broken parity model which has an economy of new parameters
beyond those of the standard model. The model predicts not only
mini-charged particles but ones which also have ``mini-mass''.
The new physics in the model occurs at low energies. The new
particles are weakly coupled to ordinary matter, but could
be discovered by low energy experimental searches for light
mini-charged particles.


\begin{thebibliography}{99}
\bibitem{par} In the literature there seems to be some confusion
as to the number of free continuous
parameters in the standard model. We find that there
are 20 parameters: 9 fermion masses, 3 KM angles, 1 KM phase, 1 Higgs
VEV and 1 Higgs quartic coupling, 3 gauge couplings, 1 $\theta$--QCD term
and 1 undetermined electric charge parameter.
\end{thebibliography}
\end{document}